# Condition Driven Adaptive Music Generation for Computer Games

Alamgir Naushad
Faculty of Computer Science and Engineering, GIK Institute, Pakistan

Tufail Muhammad
Faculty of Computer Science and Engineering, GIK Institute, Pakistan

## ABSTRACT

The video game industry has grown to a multibillion dollar, worldwide industry. The background music tends adaptively in reference to the specific game content during the game length of the play. Adaptive music should be further explored by looking at the particular condition in the game; such condition is driven by generating a specific music in the background which best fits in with the active game content throughout the length of the gameplay. This research paper outlines the use of condition driven adaptive music generation for audio and video to dynamically incorporate adaptively.

## 1. INTRODUCTION

While talking about the adaptive music, this fact basically highlight the future events which are going to occur in the upcoming events, the main emphasis of today's world research critically evaluate background music of an audio or video game to adjust or varies accordingly with the particular piece or segment of music, keeping in view the player mood to react instantly with such segment of music depending upon the visual aspect of game content through experience driven procedure content generation (EDPCG) [1]. In order to bring adaptive behavior in the background music first of all the focus is on generic approach by defining a performance matrix which should evaluate the entertainment factor by modeling all the parameters which are critical to generate an adaptive music incorporates during an active segment of current phase of the game play element and the process continues until model all the segments of the game play element. Certain techniques have been described for increasing the entertainment factors, which can be considered as the mandatory future aspect in the field of game development [2]. For generating an adaptive music (one of the element of a particular game content) by focusing the desired target upon biologically inspired evolutionary algorithms and other meta-heuristic for automatic generation of game content. The search based procedural content generation (SBPCG) can be one of the platforms to achieve the above goal, by evaluating the fitness function of the candidate solution (candidate content) using a special case of generate and test approaches [3]. The other technique which can fallow for guiding the background music to be adaptive at times whenever necessary can be done by incorporating tonal music. The tonal music can be design by constructing a network of chords using some of the network topologies. This step has to go through an experimentation phase where it can check for star, ring, mesh and other topologies to find out which best approximates with the proposed problem, the next step will introduce learning (which will reward as rank the acceptable chords using one of the network topology depending upon the dynamics of a particular game) for learning best motion in progress [4]. The design space or search space for the procedural content generation of games can be explicitly represented using domain independent procedure as parameters for game content, as the task is to generating the mood varying adaptive background music so in the design space can explicitly highlight the domain independent procedures for generating particular feature of the game content (game music) [5]. In designing and modeling adaptive music generation that varies dynamically with the corresponding event, player satisfaction is the most important parameter, for this to accomplish it must have the higher entertainment value. The approach it fallow is gradient ascent technique. Experimental results show the robustness and efficiency of this technique when practically applied to the input [6]. Automatic Sequence generation for Music that should satisfy certain arbitrary criteria, which can vary the input criteria (driven inputs) to validate Results, for such reason it should produce a system with an automatic Music generation [7].

This automatic approach of Music generation can be incorporated in to the personalized music sports video generation, the two important issues to be classified are semantic sports video content extraction and the automatic music video composition, these issues can be addressed using Multimodal cues [8].

## 2. RELATED WORK

It is sometimes most obvious to Evaluate the Accuracy of evaluation functions is one of the critical factor in computer game players, Evaluation functions Normally points out the game positions and then may be constructed manually as a weighted linear combinations of evaluation features [9]. Keeping all the important paradigms regarding music it is also important to keep focus on the Music thumbnails based on content analysis for automatic generation of music which can be used for efficient management and computation of digital Music [10]. Game engine concept also plays a crucial role in the game development, so it is indeed a very creative feature to add game engines for musical and / or sonic purposes [11]. Let us introduce a new approach if let suppose the game contents to be constantly and automatically renewed. Through this way the players can be engaged for a long duration of time; this can be done by adopting an algorithm called content generating Neuro Evolution of Augmenting Topologies [12].

For testing purpose the chosen game is super Mario AI championship, the desired output is to achieve high score (good score), for that to acquire the targeted approach certain variable strategies. To investigate the properties of benchmark, the theoretical consideration on how to find out the neural network based controllers that can be evolved using different network architecture and input spaces [13, 19-23]. For Case study and experimental purposes we can target the automatic generation of game elements (i.e. Puzzles), by using the approach of evolutionary algorithm that must optimize the puzzle to a specified level of difficulty using the fitness function [14]. Now extend the above approach [14] of an evolutionary algorithm to the Neural Networks by developing the fitness evaluator; it is an automatic process which won't require any human interaction during the evolution phase [15]. Highlighting the parameter or mapping a unique parameter to some other parameter investigates the level design parameters of platform games, like individual playing characteristics and player experience. Similarly extending this approach to super Mario AI Championship by modeling the component parameters including fun, frustration and Challenge.

A Neural Network model can be trained to map the level design parameters which shows good approximation of results and automatically enhances the player experience [16].A hybrid of video and outdoor games addresses two major issues, the first one is health related concerns and second one is the entertaining factor (which is the main area of interest), the solution to the mention problem is tangible tiles with the capability of interacting with user and each other. Genetic algorithm can also be applied to create interesting games for the mention platform [17]. Computer games prove to be the great source of entertainment for all age groups, so from the composer point of view it's a tedious job to keep the common interest of the people of all age groups to influence entertainment value, which is very subjective, so the factors which greatly influence the entertainment value are the type of game and the contents of the game, so the keen interest is to study the quantitative measures for entertainment factor [18].

By passing concluding remarks on the literature survey we have come across some of the common similarities among all the referred literature work. First of all the desired output in the entire literature work focuses on the dynamic or adaptive music generation by applying different techniques or approaches, it is important to mention here that most of the literature work survey visited was either considering the entire game or some particular event to be modeled dynamically by generating adaptive music in the entire game or the particular event respectively. The approach simply considers the particular event components or some specific parameters and then calculate the impact of the independent parameter which affect mood of the player who is playing the game through his/her Game experience a different example of use of music [25,26] in the domain of data mining[24] is via sonification.

This paper investigates how the adaptive music can be applicable for particular dataset of a game .Dataset can be composed of having unique player attributes or parameter and then generating the dynamic music generation on the basic of certain conditions(flexible), imposed upon every attribute and then assigning the net effect of all the attributes to a particular class label (classifying a particular type of music) , which in other words should model the current segment of game play in terms of music.

## 3. RANDOM MUSIC GENERATION

For adaptive random music generation, first of let us create random matrix array in some range, as this matrix will comprise upon random numbers which explicitly represents a wide range of possibility as an input filter for the music generation .sound can be generated by taking random number into account.

We can argue that the music generated will consist upon unique segments of sound. This differentiation is due to the matter of fact that in random matrix there is a wide range of numbers and in the output music file every music segment corresponds to one such random number, hence this is the first step towards the adaptive music generation, but it is not for special kind of consideration because it is very hard to explicitly recognize a unique data pattern from such files, so in order to solve this issue, the focus is on a predefined dataset to generate an adaptive or dynamic music.

## 4. DATASET APPROACH FOR ADAPTIVE MUSIC GENERATION

In this phase let us randomly picked a predefined dataset (in which data can be easily recognized) from one of the game, the proposed dataset consists upon some unique attributes representing some parameters of a player life like energy of the player, score and level of the game. Each of the attributes will dynamically update itself depending upon the performance of the player who is playing the music which at the background (for every attribute) corresponds to some random value, which is a flexible in nature and adjusted in accordance with the player performance for every attribute. Let us say if the player life is greater than the random value map to this attribute will be a large scale value and vice versa. Same is true for all attributes by dynamically updating itself during the entire game length of play.

Now in order to accomplish the task of adaptive music for selected dataset let define another attribute by the name class label. In the class label we have a verity of sounds each sound corresponds to some unique music. Let $(\alpha)$ represent sad music, $(\beta)$ which refers happy music, $(\gamma)$ for normal music and $(\varphi)$ for angry music. All these different categories of music will execute depending upon the conditional criteria $(\forall)$ to be fulfilled, the conditional criteria for all music categories' are given as follows.

$$\forall = \begin{cases} \alpha & if\ \tau \leq 1 \delta\delta\sigma \leq 20 \delta\delta p < 2000 \delta\delta\mu \leq 2 \\ \beta & if\ \tau \geq 4 \delta\delta\sigma \geq 80 \delta\delta p < 8000 \delta\delta\mu \geq 8 \\ \gamma & if\ \tau \leq 3 \delta\delta\sigma \leq 50 \delta\delta p < 5000 \delta\delta\mu < 6 \\ \varphi & Else \end{cases}$$

Where $\tau$ represent player life, $\sigma$ is for energy of player, $p$ for score, $\mu$ for level of the game.

Referring to the block diagram one can generate random dataset matrix of 5x4, after this we compare the data set with different conditions i.e. row one is compared with four conditions in case row one matches with any of these condition according to that condition particular music will be played, by extending the above approach for the reaming rows as well.

## 5. EXPERIMENTAL RESULTS

First of all let randomly generate different attributes (player life, energy of player, score, and level) for different players; the next random value is compare with the conditional criteria when the condition meets and so on.

Let's say $\beta$ ,then $\beta$ kind of music will be generated, then for other random dataset the condition repeats back thus introducing adaptive or dynamic music generation. During the game play let us experimentally performed some of the data sets which are mentioned in appendix-I.

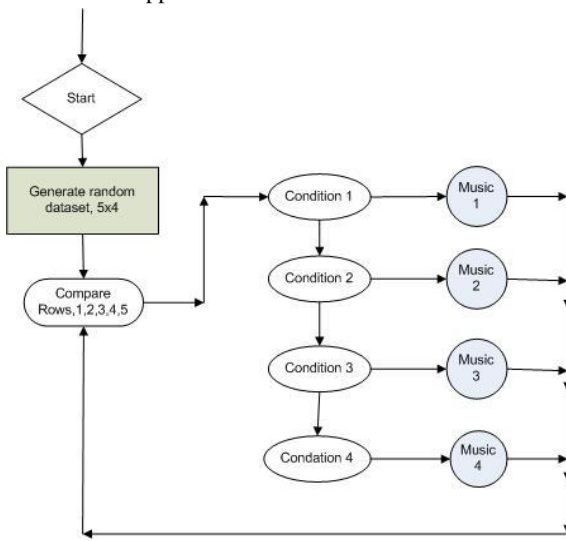

**Fig. 1 Block diagram for Adaptive music Generation**

In the dataset(random) music has been generated according to mention condition 1st ,3rd ,4th and 5th row fulfill the condition for angry music, so angry music will be executed, and the 2nd row satisfies normal music so normal music has been played. In the 2nd dataset all the four rows satisfied condition for angry music, so angry music is played, only 5th row fulfill for normal music, so Normal music will be played. This dataset satisfies only the condition for angry music. The resultant music values which are generated for the selected dataset are

Normal =0.1100  0.1110  0.1140  0.1090  0.0970  0.1080

Anger  = 0.0970  0.1100  0.1030  0.1010  0.1140  0.0320

Happy  = 0.1040  0.0 970  0.1120  0.1120  0.1210  0.0320

Sad    = 0.1150  0.0 970  0. 1000  0.0320  0.0320  0.0320

**User Survey:**

The user survey is performed by comparing and verify the proposed adaptive music generation against static music generation using some traditional and conventional methods. The data set was tested for several number of professional users. The users were asked to rank both the static and dynamic approaches of music generation on the basis of the following criteria: 1→Represent dislike,  2→ Represent Normal and 3→Like. Table 1 summarizes the suer survey results.

**Table 1. User Survey Results**

| Subject | Age | Gender | Profession | Adaptive Music Generation | Static Music |
|---|---|---|---|---|---|
| 1 | 24 | M | Student | 3 | 1 |
| 2 | 40 | M | Business Man | 2 | 3 |
| 3 | 30 | F | Doctor | 3 | 2 |
| 4 | 35 | M | Lawyer | 3 | 1 |
| 5 | 45 | F | Journalist | 1 | 1 |
| 6 | 30 | M | Engineer | 3 | 2 |
| 7 | 50 | M | Teacher | 3 | 1 |

The User Survey data (Table 1) is represented in the graphical form using figure 2 . the results suggest that 71% of the subjects have liked the most those contents of music generated via adoptive music generation.

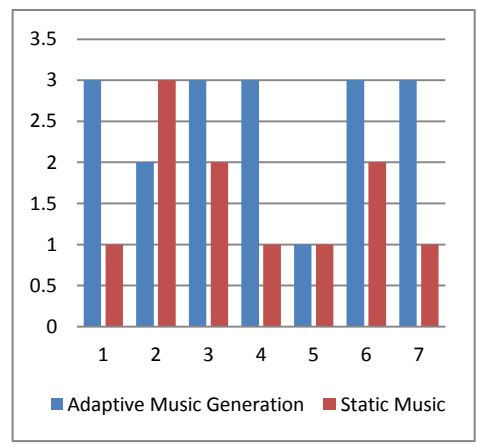

**Figure 2. User Ratings.**

## 6. CONCLUSIONS

From the Experimental result it is obvious that the final Result of the datasets varies dynamically with randomly generated Numerical data elaborating versatility in a sense by generating Adaptive Music. Secondly it can be verified the conditional criteria for all the mention parameters of the dataset which updates itself by matching the given Condition. Adaptive behavior can be noticed in the final output which lists all the mention categories for Unique Music. One can easily see the dynamic behavior for the corresponding event by modeling the unique components (which is a collection of unique attributes or parameters), representing the adaptive nature of an event in terms of music by highlighting the unique parameters or attributes. Randomness in the output music file represents that the mood of the player is driven on the basis of certain condition which is attribute or parameter driven, thus it is a perfect platform to keep the entertainment value in terms of dynamic behavior.

# APPENDIX-I

1st Dataset result:

| $\tau$ | $\sigma$ | $p$ | $\mu$ | $\forall$ | | | | | |
|--------|----------|--------|--------|--------|--------|--------|--------|--------|--------|
| 0.0020 | 0.0660 | 7.2711 | 0.0070 | 0.0970 | 0.1100 | 0.1030 | 0.1010 | 0.1140 | 0.0320 |
| 0.0036 | 0.0342 | 3.0929 | 0.0055 | 0.1100 | 0.1110 | 0.1140 | 0.1090 | 0.0970 | 0.1080 |
| 0.0033 | 0.0290 | 8.3850 | 0.0044 | 0.0970 | 0.1100 | 0.1030 | 0.1010 | 0.1140 | 0.0320 |
| 0.0026 | 0.0341 | 5.6807 | 0.0069 | 0.0970 | 0.1100 | 0.1030 | 0.1010 | 0.1140 | 0.0320 |
| 0.0033 | 0.0534 | 3.7041 | 0.0062 | 0.0970 | 0.1100 | 0.1030 | 0.1010 | 0.1140 | 0.0320 |

2nd Dataset result:

| $\tau$ | $\sigma$ | $p$ | $\mu$ | $\forall$ | | | | | |
|--------|----------|--------|--------|--------|--------|--------|--------|--------|--------|
| 0.0034 | 0.0503 | 1.9343 | 0.0070 | 0.0970 | 0.1100 | 0.1030 | 0.1010 | 0.1140 | 0.0320 |
| 0.0001 | 0.0709 | 6.8222 | 0.0038 | 0.0970 | 0.1100 | 0.1030 | 0.1010 | 0.1140 | 0.0320 |
| 0.0027 | 0.0429 | 3.0276 | 0.0086 | 0.0970 | 0.1100 | 0.1030 | 0.1010 | 0.1140 | 0.0320 |
| 0.0015 | 0.0305 | 5.4167 | 0.0085 | 0.0970 | 0.1100 | 0.1030 | 0.1010 | 0.1140 | 0.0320 |
| 0.0033 | 0.0190 | 1.5087 | 0.0059 | 0.1100 | 0.1110 | 0.1140 | 0.1090 | 0.0970 | 0.1080 |

3rd Dataset result:

| $\tau$ | $\sigma$ | $p$ | $\mu$ | $\forall$ | | | | | |
|--------|----------|--------|--------|--------|--------|--------|--------|--------|--------|
| 0.0038 | 0.0762 | 6.1543 | 0.0041 | 0.0970 | 0.1100 | 0.1030 | 0.1010 | 0.1140 | 0.0320 |
| 0.0009 | 0.0456 | 7.9194 | 0.0094 | 0.0970 | 0.1100 | 0.1030 | 0.1010 | 0.1140 | 0.0320 |
| 0.0024 | 0.0019 | 9.2181 | 0.0092 | 0.0970 | 0.1100 | 0.1030 | 0.1010 | 0.1140 | 0.0320 |
| 0.0019 | 0.0821 | 7.3821 | 0.0041 | 0.0970 | 0.1100 | 0.1030 | 0.1010 | 0.1140 | 0.0320 |
| 0.0036 | 0.0445 | 1.7627 | 0.0089 | 0.0970 | 0.1100 | 0.1030 | 0.1010 | 0.1140 | 0.0320 |